\documentclass[a4paper]{article}
\usepackage{epsfig,amsmath,amsfonts,amssymb,setspace,multirow,textcomp}
\usepackage[T1]{fontenc}
\makeatletter
\renewcommand{\@oddhead}{\textit{Advances in Astronomy and Space Physics} \hfil}
\renewcommand{\@evenfoot}{\hfil \thepage \hfil}
\renewcommand{\@oddfoot}{\hfil \thepage \hfil}
\makeatother

\renewenvironment{thebibliography}[1]{\begin{oldthebibliography}{#1}\setlength{\parskip}{0ex}\setlength{\itemsep}{0ex}}{\end{oldthebibliography}}

\begin{document}
\fontsize{11}{11}\selectfont 
\title{Dynamics of dark energy in collapsing halo of dark matter}
\author{\textsl{M.~Tsizh}, B.~Novosyadlyj, }
\date{\vspace*{-6ex}}
\maketitle
\begin{center} {\small Ivan Franko National University of Lviv, Kyryla and Methodia str., 8, Lviv, 79005, Ukraine\\}
\end{center}

\begin{abstract}
We investigate the non-linear evolution of spherical density and velocity perturbations of dark matter and dark energy in the expanding Universe. For that we have used the conservation and Einstein equations to describe the evolution of gravitationally coupled inhomogeneities of dark matter, dark energy and radiation from linear stage in the early Universe to the non-linear one at the current epoch. The simple method of the numerical integration of the system of non-linear differential equations for evolution of the central part of halo is proposed. The results are presented for halo of cluster ($k=2$ Mpc$^{-1}$) and supercluster scales ($k=0.2$ Mpc$^{-1}$) and show that quintessential scalar field dark energy with small value of effective speed of sound $c_s<0.1$ can give noticeable impact on the formation of large scale structures in the expanding Universe. \\[1ex]
\end{abstract}

\section*{\sc introduction}
\indent \indent Dark energy is mysterious dark component responsible for accelerated expansion of the Universe became an object of numerous studies in the last two decades. There are many possible explanations \cite{euclid}, \cite{novosbook} of the nature of dark energy. One of the most promising among them is scalar field dark energy, which can be modelled as perfect fluid. In fact, having just a few parameters -  the density $\rho_{de} = \rho_{cr}\Omega_{de}$, the equation of state (EoS) parameter $w_{de}=\rho/p$ and effective speed of sound $c_s^2=\delta p/\delta \rho$, this model can perfectly explain latest cosmological observational data (\cite{serg1}, \cite{serg2}), keeping a large volume of parameters space of possible values. Therefore, it would be interesting to study behaviour of perturbations of such dark energy on non-cosmological scales. This may yield more constrains on parameters of the model.

It was already shown \cite{arbey}, \cite{wett} that non-linear perturbations of scalar field dark energy can in principle influence structure formation even on galaxy scales. Finally, in the last decade there was a number of papers (see \cite{babich} for review), that studied accretion of dark energy on compact objects (black holes mainly) in hydrodynamical approach. In our works \cite{tsizh1,tsizh2}  we have shown that 
nevertheless such stationary accretion does not change the gravitating mass of the central object, in principle, it can give detectable contribution to  dynamics of bodies in its vicinity, which would be possible to extract from current observations.

Here we analyse the dynamics of dark energy in the collapsing halo of dark matter. In the next chapter we present the system of equations, which we use to solve the problem of evolution of spherical scalar perturbations in 3-component expanding Universe from early linear stage to highly non-linear one, when dark matter halo forms. In the last chapter we present numerical solutions and discuss their main features.

\section*{\sc equations for evolution of perturbations}
\indent  
We suppose that the Universe is spatially flat and filled by matter ($m$), dark energy ($de$) and radiation ($r$), the metric of the background space-time is
Friedmann-Robertson-Walker (FRW) one. Each component is described in the perfect fluid approximation by energy density $\varepsilon$, pressure $p$ and four-velocity $u^i$. The equation of state for each component can be presented as $p=w\varepsilon$, with $w_m=0$ for matter, $w_r=1/3$ for radiation and $w_{de}<-1/3$ for dark energy. We assume that dark energy is scalar field with $w_{de}=const$. 
The goal of the paper is analysis the evolution of spherical halo from the linear stage in the early radiation dominated epoch, through the turnaround point to highly non-linear stage, infall of matter before virialization, at dark energy dominated epoch.
The local spherical perturbation distorts the FRW metric, so that it becomes
\begin{equation}
ds^2=e^{\nu(t,r)}dt^2-a^2(t)e^{\mu(t,r)}[dr^2+r^2(d\theta^2+\sin^2\theta d\varphi^2)],\label{ds_sph}
\end{equation}
where the metric functions $\nu(t,r)$ and $\mu(t,r)$ vanish at cosmological background.
At the linear stage, when $\nu(t,r)\sim\mu(t,r)\ll1$ the metric (\ref{ds_sph}) becomes the metric of conformal-Newtonian (longitudinal) gauge \cite{Bardeen1980} in spherical coordinates, since  $e^{\nu(t,r)}\approx1+\nu$, $e^{\mu(t,r)}\approx1+\mu$.
We rewrite the energy-momentum tensor of dark energy, $T_{i\,(de)}^k = (\varepsilon_{de}+p_{de})u_{i\,(de)}u^k_{de} - \delta_i^kp_{de}$, in terms of proper 3-velocity of fluid $v_{de}$ (measured in local frame), which has only radial component. The relation between components of 4-velocity    $u^i_{de}$ and 3-velocity $v_{de}$ of fluid are as follows
\begin{equation}
u_{i\,(de)} = \left\{\frac{e^{\nu/2}}{\sqrt{1-v_{de}^2}},-\frac{av_{de}e^{\mu/2}}{\sqrt{1-v_{de}^2}},0,0\right\},\quad 
u^{i}_{de}= \left\{\frac{e^{-\nu/2}}{\sqrt{1-v_{de}^2}},\frac{a^{-1}v_{de}e^{-\mu/2}}{\sqrt{1-v_{de}^2}},0,0\right\},
\label{u-v}
\end{equation}
where $v_{de}$ is in the units of speed of light. The non-zero components of energy-momentum tensor are
\begin{equation}
 T_0^0 = \frac{\varepsilon_{de}+v_{de}^2p_{de}}{1-v_{de}^2},\quad
 T_0^1 = \frac{\varepsilon_{de}+p_{de}}{1-v_{de}^2}a^{-1}v_{de}e^{(\nu-\mu)/2},\quad
 T_1^1 =-\frac{v_{de}^2\varepsilon_{de}+p_{de}}{1-v_{de}^2},\quad
 T_2^2 = T_3^3 = - p_{de}.
\label{tei}
\end{equation}
We decompose both density and pressure on background averaged and perturbed parts as $\varepsilon_{de}=\bar{\varepsilon}_{de}(1+\delta_{de})$, $p_{de}=\bar{\varepsilon}_{de}(w_{de}+c_s^2\delta_{de}-3\dot{a}(1+w_{de})(c_s^2-w_{de})\int{e^{(\mu-\nu)/2}v_{de}dr})$, where density perturbation of each component is defined as $\delta(t,r)\equiv(\varepsilon(t,r)-\bar{\varepsilon}(t))/\bar{\varepsilon}(t)$. We study the model of dark energy, for which both $w_{de}$ and $c_s^2$ are constant. The integral term in the pressure decomposition comes from non-adiabatic part of pressure perturbation of scalar field dark energy (details can be found in papers \cite{Hu1998,Hu2004,Novosyadlyj2015}). The presence of this term makes our equation of state non-barotropic, as we work in frame different from proper dark energy one and hence, relation $p_{de}=w_{de}\rho_{de}$ holds only for averaged parts of this component. Taking into account this decomposition, and keeping terms with $v_{de}^0$, $v_{de}^1$ and $v_{de}^2$ only, we 
obtain the following energy-momentum tensor components of dark energy:
\begin{eqnarray}
&&T_0^0 =\bar{\varepsilon}_{de}(1+\delta_{de})+\bar{\varepsilon}_{de}(1+w_{de}+(1+c_s^2)\delta_{de})v_{de}^2, \nonumber \\
&&T_0^1 =\bar{\varepsilon}_{de}\left[1+w_{de}+(1+c_s^2)\delta_{de}-3\dot{a}(1+w_{de})(c_s^2-w_{de})\int{e^{(\mu-\nu)/2}v_{de}dr}\right]a^{-1}v_{de}e^{(\nu-\mu)/2}, \nonumber \\
&&T_1^0 =-\bar{\varepsilon}_{de}\left[1+w_{de}+(1+c_s^2)\delta_{de}-3\dot{a}(1+w_{de})(c_s^2-w_{de})\int{e^{(\mu-\nu)/2}v_{de}dr}\right]a v_{de}e^{(\mu-\nu)/2}, \label{tei2}\\
&&T_1^1 = - \bar{\varepsilon}_{de} \left(w_{de} + c_s^2\delta_{de} -3\dot{a}(1+w_{de})(c_s^2-w_{de})\int{e^{(\mu-\nu)/2}v_{de}dr}\right)-\bar{\varepsilon}_{de}\left(1+w_{de}+(1+c_s^2)\delta_{de}\right)v_{de}^2,\nonumber\\
&&T_2^2 = T_3^3 =-\bar{\varepsilon}_{de}\left[w_{de}+c_s^2\delta_{de}
-3\dot{a}(1+w_{de})(c_s^2-w_{de})\int{e^{(\mu-\nu)/2}v_{de}dr}\right]. \nonumber
\end{eqnarray}
 Hereafter, one can get the corresponding equations and expressions for dark matter and radiation components from equations and expressions for dark energy  just by putting $w_{de}=c_s^2=0$ for matter and $w_{de}=c_s^2=1/3$ for radiation.

To find the evolution of density and velocity perturbations we use two conservation equations, which have general covariant form
$$T^k_{i;k}=\frac{1}{\sqrt{-g}}\frac{\partial(\sqrt{-g}T^k_i)}{\partial x^k}-
\frac{1}{2}\frac{\partial g_{kk}}{\partial x^i}g^{kk}T^k_k=0 $$
(for $i=0$ it is continuity equation, for $i=1$ it is motion equation).
Substituting our energy-momentum tensor (\ref{tei2}) one obtains:
\begin{eqnarray}
&&\dot{\delta}_{de}(1+(1+c_s^2)\delta_{de}v^2) + 3\frac{\dot{a}}{a} \delta_{de} (c_s^2-w_{de}) + \left[1+w_{de}+(1+c_s^2)\delta_{de}\right]\left(\frac{3\dot{\mu}}{2}+(1-3w_{de}+2\dot{\mu})v^2_{de}+2\dot{v}_{de}v_{de}\right)+\nonumber \\
&&+a^{-1}e^{(\nu-\mu)/2}\left[(1+c_s^2)\delta'_{de} v_{de} + \left(1+w_{de}+(1+c_s^2)\delta_{de} \right)\left(v_{de}(\nu' + \mu' + \frac{2}{r}) + v'_{de}\right)\right]+ \label{de_cl0} \\
&&+3\dot{a}(1+w_{de})(c_s^2-w_{de})\left[\left(\frac{3\dot{a}}{a}+\frac{3\dot{\mu}}{2} + a^{-1}e^{(\nu-\mu)/2}\left(v_{de}(\nu' + \mu' + \frac{2}{r}) + v'\right)\right)\int e^{(\mu-\nu)/2}v_{de}dr - v^2_{de}  \right] = 0 \nonumber
\end{eqnarray}
\begin{eqnarray}
&&\dot{v}_{de} + v_{de}\left(\frac{\dot{a}}{a}(1-3c_s^2) + 2\dot{\mu}\right) +  \frac{\delta_{de}'e^{(\nu-\mu)/2}}{a(1+w_{de})}\left(c_s^2 + \frac{1+c_s^2}{1+w_{de}}v^2_{de}\right) +\nonumber\\
&&\frac{e^{(\nu-\mu)/2}}{a}\left(1+\delta_{de}\frac{1+c_s^2}{1+w_{de}}\right)\left(\frac{\nu'}{2} + (\nu'+\mu')v_{de}^2 + 2v_{de}(v_{de}' + \frac{v_{de}}{r}) \right) + \frac{1+c_s^2}{1+w_{de}} \left(\dot{\delta}_{de}v + \dot{v}_{de}\delta + 2\dot{\mu}v_{de}\delta_{de}+\right.\nonumber\\
&&\left.(1-3w_{de})\frac{\dot{a}}{a}v_{de}\delta_{de}  \right)
-3\frac{\dot{a}}{a}(c_s^2-w_{de})\left[\left(\dot{v}_{de}+(1-3w_{de})\frac{\dot{a}}{a}v_{de}+2\dot{\mu}v_{de}+
\frac{\nu'}{2a}e^{(\nu-\mu)/2}\right)\int{e^{(\nu-\mu)/2}v_{de}dr}+ \right.\nonumber\\ &&\left.v_{de}\int{e^{(\nu-\mu)/2}\left(\dot{v}_{de}+\frac{\ddot{a}}{\dot{a}}v_{de}+\frac{\dot{\mu}-\dot{\nu}}{2}v_{de}\right)dr}\right]=0.  \label{de_cl1}
\end{eqnarray}
The continuity equation for background density, which we also use, is $\dot{\bar{\varepsilon}}_{de}+3\frac{\dot{a}}{a}(1+w_{de})\bar{\varepsilon}_{de}=0$. 

To find the metric functions we exploit the Einstein equations
\begin{equation}
R^i_j -\frac{1}{2}\delta^i_j R = \varkappa \left(T^i_{j\;(de)}+T^i_{j\;(m)}+T^i_{j\;(r)}\right)
\nonumber
\end{equation}
If we construct the equation $G^1_1-G^2_2=\varkappa\left(T^1_1-T^2_2\right)$ then we will see that at the linear stages $\mu=-\nu$. At the non-linear stage, the right hand side of this equation equals zero at the center of perturbation too, which gives again reduction to only one potential. In this paper we  analyse the dynamics of dark matter and dark energy in the central part of spherical overdensity only, therefore we accept $\mu=-\nu$ approximation, that gives us the possibility to use only one $(^0_0)$ Einstein equation for determination of one metric function $\nu(t,r)$: 
\begin{eqnarray}
&&-3\frac{\dot{a}}{a}\dot\nu+\frac{3\dot{\nu}^2}{4}+3\frac{\dot{a}^2}{a^2}\left(1-e^{\nu}\right)+
\frac{1}{a^2}\left[\nu''+\frac{2}{r}\nu'+\frac{\nu'^2}{4}\right]=3\frac{\dot{a}^2}{a^2}e^{\nu}
\frac{\Omega_ra^{-1}\delta_r+\Omega_{de}a^{-3w_{de}}\delta_{de}+\Omega_m\delta_m}{\Omega_m+\Omega_ra^{-1}+\Omega_{de}a^{-3w_{de}}}.
\label{ee00}
\end{eqnarray}
Here we used the notations
$\Omega_N\equiv\bar{\varepsilon}^0_{N}/(\bar{\varepsilon}^0_{m}+\bar{\varepsilon}^0_{de}+\bar{\varepsilon}^0_{r})$, where $N$ denotes the type of fluid and ``0'' marks the value at current epoch. In the flat three component universe $\Omega_m+\Omega_r+\Omega_{de}=1$.
Eqs. (\ref{de_cl0})-(\ref{ee00}) have non-relativistic limit (Appendix), which in case of dark matter ($c_s^2=w_{de}=0$) coincide with well known classical hydrodynamic equations and Poisson equation accordingly.

We are interested in the cluster and supercluster scales of perturbations, for which $\nu\ll1$, $v_m\ll1$ in the bulk of object, while $\delta_m\gg1$ in their halos. So, the equations can be essentially reduced by neglecting of the terms like, $\nu^2$, $\nu v_m$, $\nu v_{de}$ and higher order terms. The terms with $v\nabla{v}$ must be kept, since they are important at highly non-linear stage. The final reduced form of these equations, which we use in the code, can be found in our upcoming paper \cite{Novosyadlyj2015}. 

Therefore, we have seven 1st-order partial differential equations for seven unknown functions  
\begin{equation}
\delta_m(t,r), \quad  v_m(t,r), \quad  \delta_{de}(t,r), \quad  v_{de}(t,r), \quad  \delta_{r}(t,r), \quad  v_{r}(t,r), \quad  \nu(t,r),
\label{functions}
\end{equation}
which can be solved numerically for given initial conditions. 

At the early epoch the amplitudes of cosmological perturbations of space-time metric, densities and velocities are small and the eqs. (\ref{de_cl0})-(\ref{ee00}) can be linearized for all components. Moreover, we can present each function of $(t,r)$ as product of its amplitude, which depends on $t$ only, and some function of radial coordinate $r$, which describes the initial profile of spherical perturbation, which can be expanded into series of some orthogonal functions, e.g. spherical ones in our case. In particular, we can present the perturbations of the metric, density and velocity of N-component as follows
$$\nu(t,r)=\tilde{\nu}(t)\frac{\sin{kr}}{kr}, \quad \delta_N(t,r)=\tilde{\delta}_N(t)\frac{\sin{kr}}{kr}, \quad v_N(t,r)=\tilde{v}_N(t)\left( \frac{\sin{kr}}{kr} \right)'
=\tilde{v}_N(t)k\left(\frac{\cos{kr}}{kr}-\frac{\sin{kr}}{k^2r^2}\right), $$
In the analysis of the evolution of central part of spherical halo we can decompose $r$-function in the Taylor series and keep only leading terms:
$$f_k(r)\approx 1,\quad f_k'(r)\approx -\frac{1}{3}k^2 r,\quad \int f'(r) dr \approx 1,\quad f_k''(r) + \frac{2f_k(r)}{r} \approx -k^2,$$
where $f_k(r)=\sin{kr}/kr$. It gives the possibility to reduce the system of seven partial differential equation for unknown functions
(\ref{functions}) to the system of seven ordinary differential equations for their amplitudes
 \begin{equation}
\tilde{\delta}_m(t), \quad  \tilde{v}_m(t), \quad  \tilde{\delta}_{de}(t), \quad  \tilde{v}_{de}(t), \quad  \tilde{\delta}_{r}(t), \quad  \tilde{v}_{r}(t), \quad  \tilde{\nu}(t),
\label{amplitudes}
\end{equation}
which is presented in \cite{Novosyadlyj2015}.

We set the adiabatic initial conditions in the following way. To find the relations between amplitudes at some $a_{init}\ll1$ when the scales of interested here gravitationally bound systems were essentially larger than horizon scale, $a_{init}k^{-1}\gg t$ we have used the linearization of system (\ref{de_cl0})-(\ref{ee00}), which has analytical exact solution for one component Universe. Obviously, that for cluster and supercluster scales it was early radiation-dominated epoch, when $\bar{\varepsilon}_{r}\gg\bar{\varepsilon}_{m}\gg\bar{\varepsilon}_{de}$, so, the matter and dark energy can be treated as test components. The amplitude of metric function $\tilde{\nu}$ is defined by density perturbations of the relativistic component. The non-singular solution of the corresponding equations has asymptotic values at $a_{init}$
$\tilde{\nu}^{init}=-C, \quad \tilde{\delta}^{init}_{r}=C, \quad \tilde{v}^{init}_{r}=C/[4a_{init}H(a_{init})]$,
where $C$ is some constant (see for details \cite{Novosyadlyj2007}). The solutions of equation for matter and dark energy as test components give the asymptotic values for superhorizon perturbations at $a_{init}$:
$\tilde{\delta}_{m}^{init}=3C/4,  \quad \tilde{v}_{m}^{init}=C/[4a_{init}H(a_{init})],  \quad
\tilde{\delta}_{de}^{init}=3(1+w)C/4,  \quad \tilde{v}_{de}^{init}=C/[4a_{init}H(a_{init})]. $
These relations contains only single constant $C$, which value specifies the initial amplitudes of perturbations in all components. 
Below we put $C=2.6\cdot10^{-5}$ for cluster scales $k=2$ Mpc$^{-1}$ and $C=6.5\cdot10^{-5}$ for supercluster scales $k=0.2$ Mpc$^{-1}$ at time when $a=10^{-10}$.  
\begin{figure*}
\includegraphics[width=0.33\textwidth]{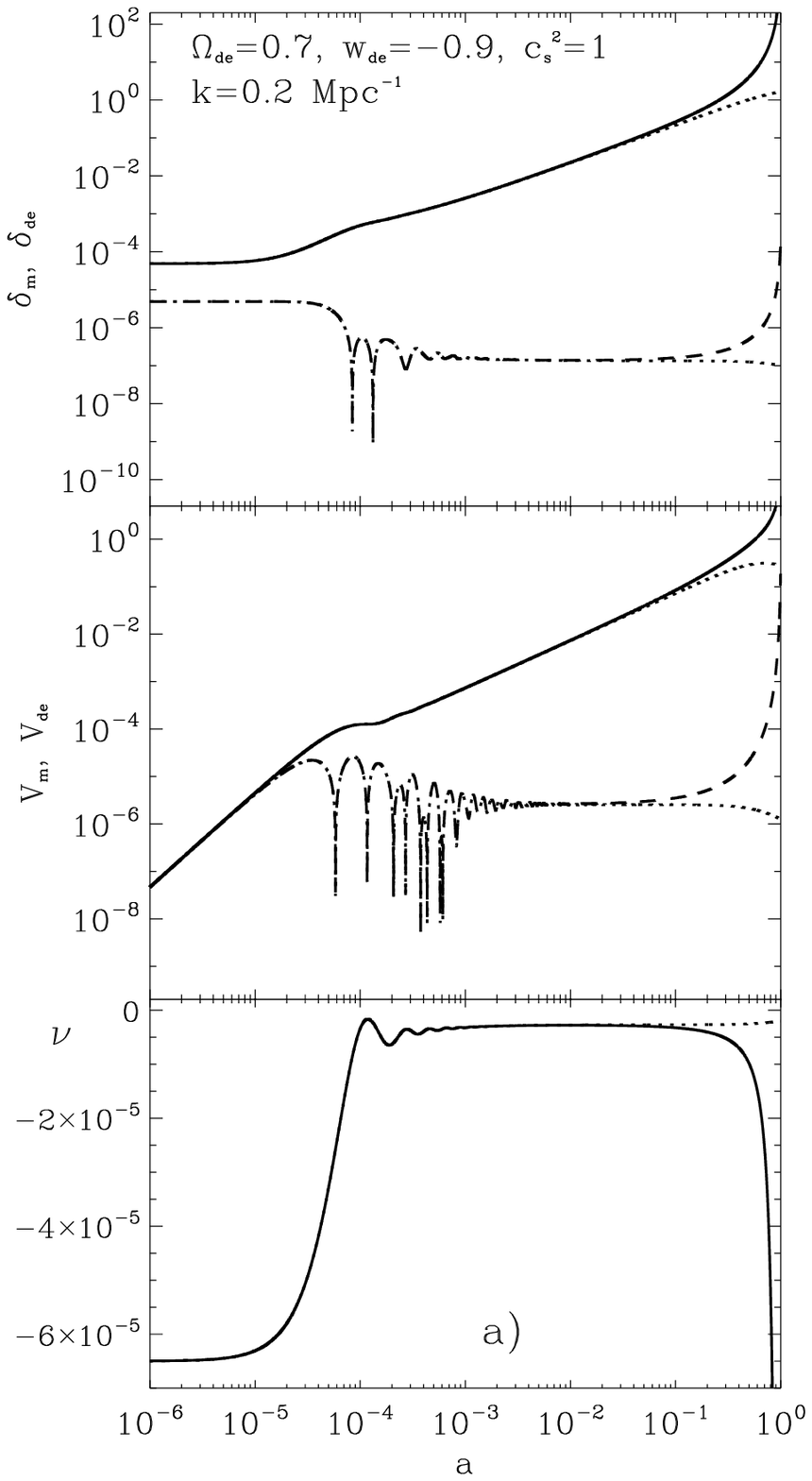}
\includegraphics[width=0.33\textwidth]{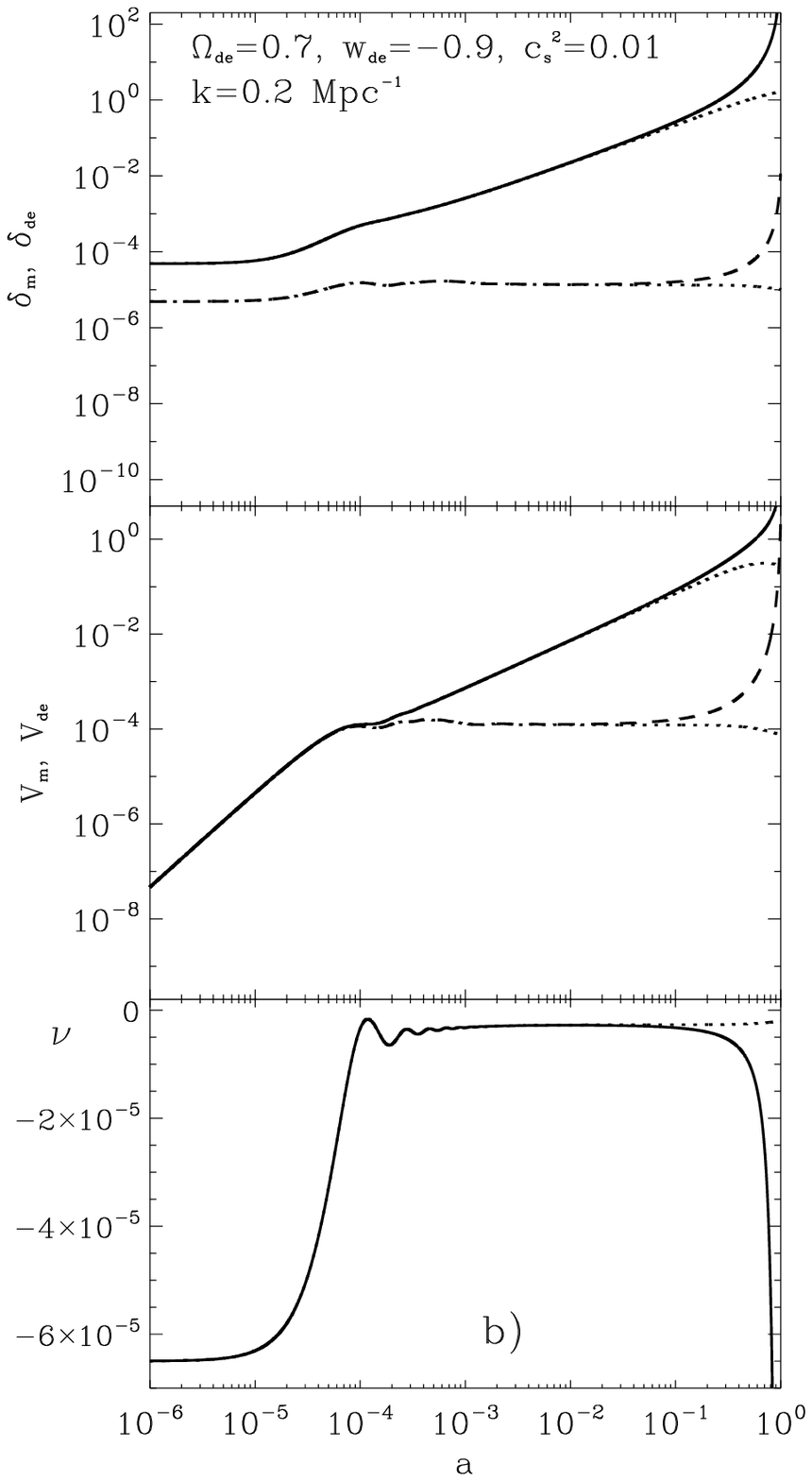}
\includegraphics[width=0.33\textwidth]{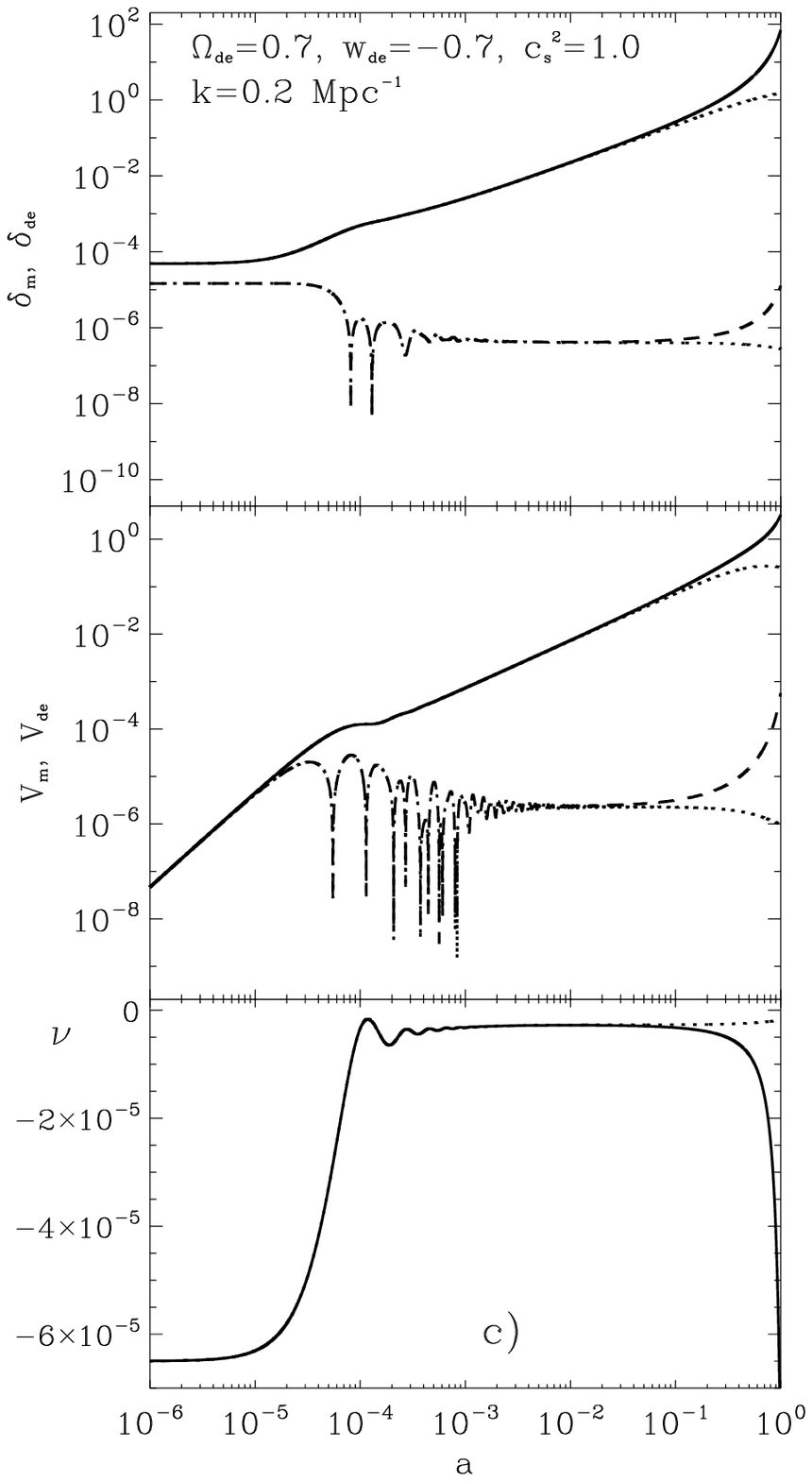}
\includegraphics[width=0.33\textwidth]{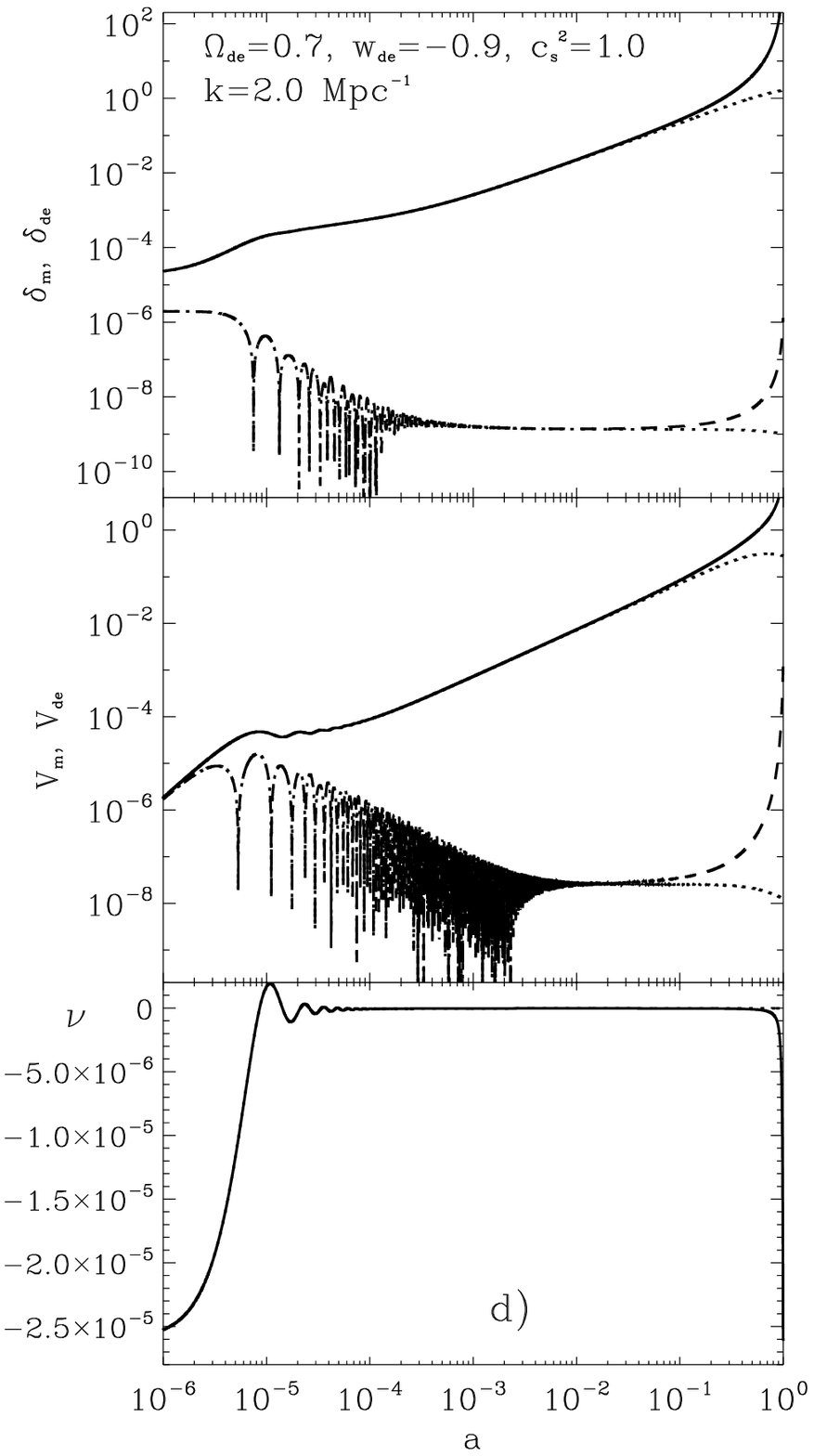}
\includegraphics[width=0.33\textwidth]{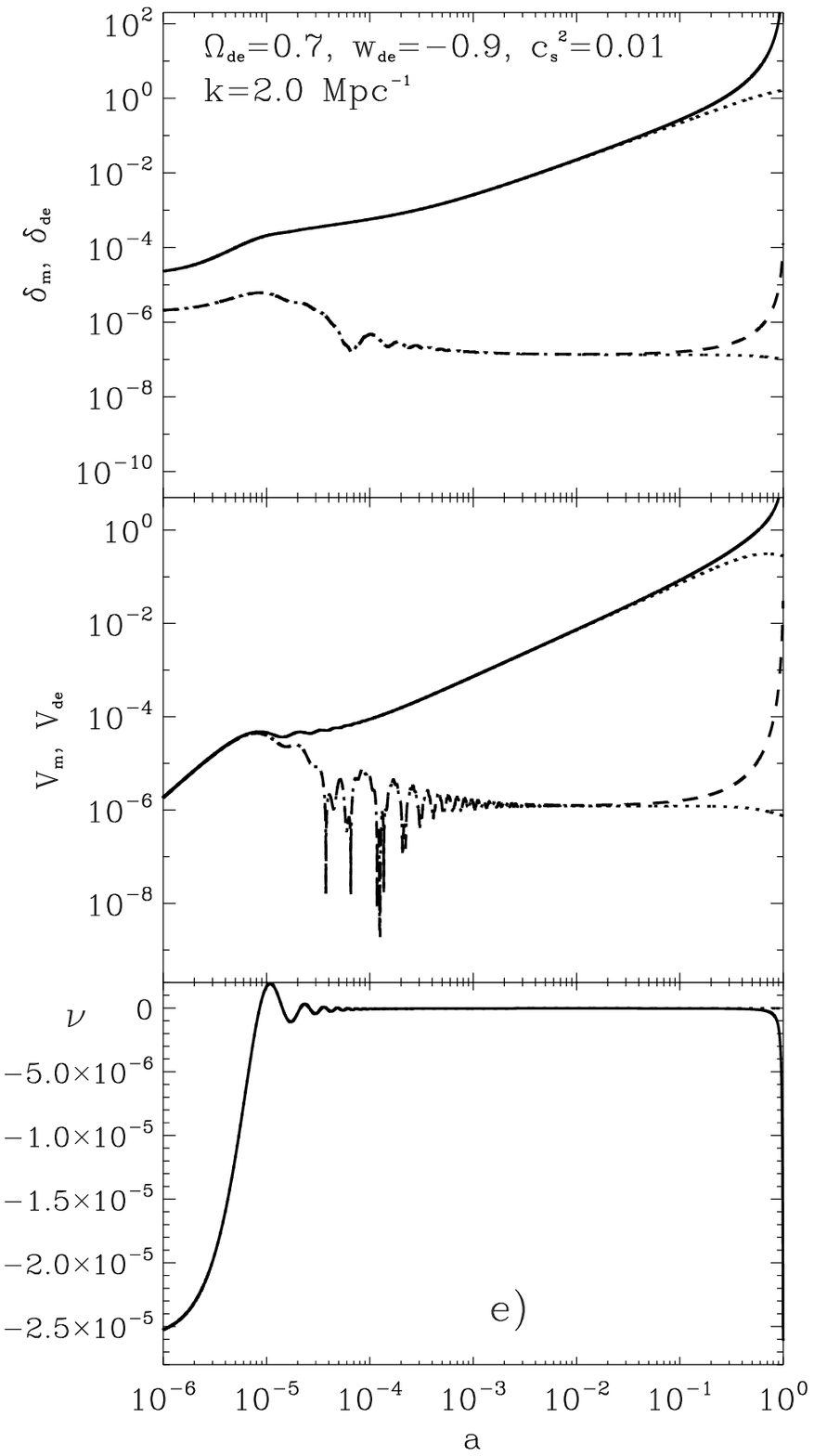}
\includegraphics[width=0.33\textwidth]{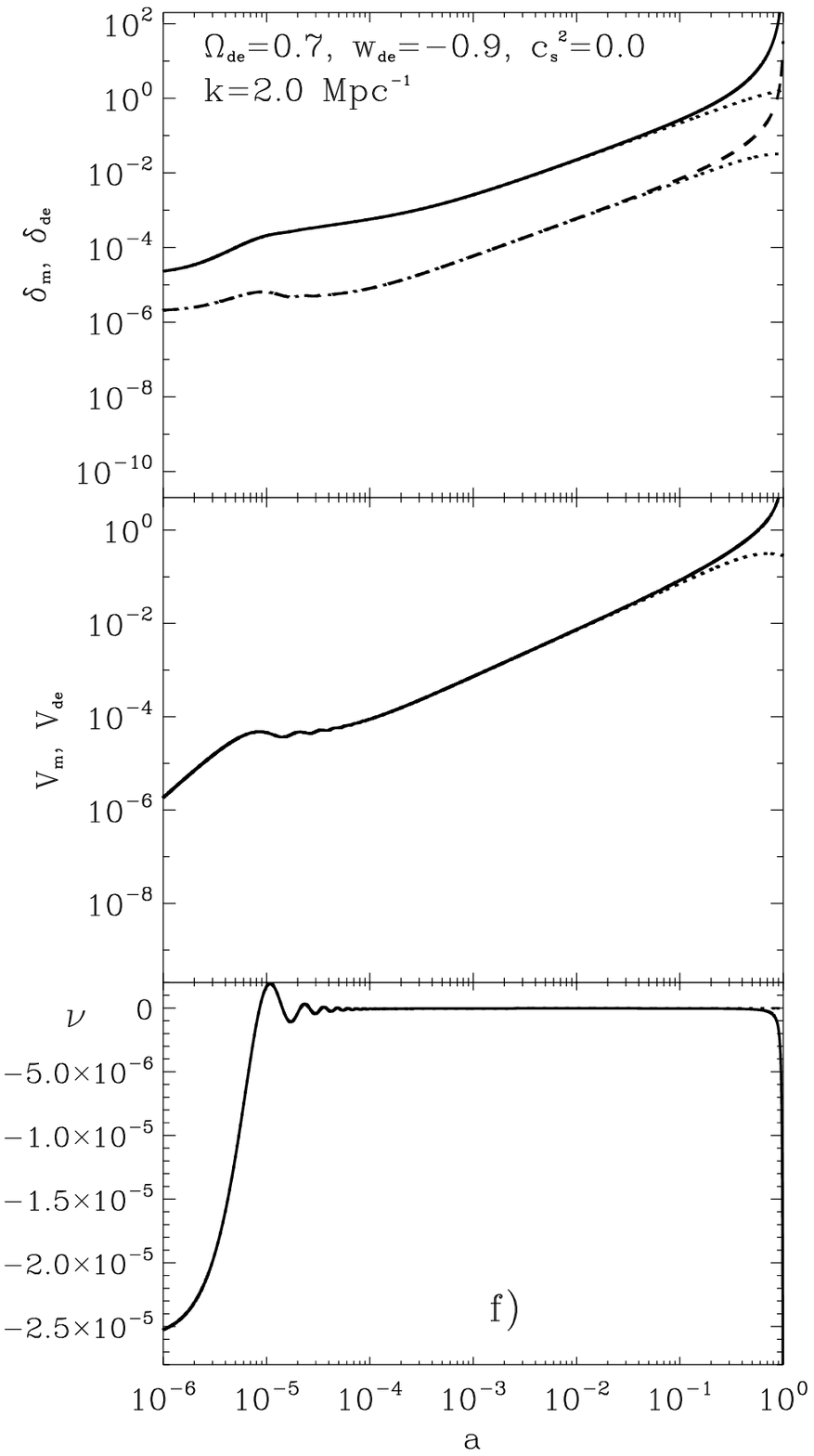}
\caption{Evolution of the amplitudes of density perturbations of matter $\tilde{\delta}_m$ and dark energy $\tilde{\delta}_{de}$ (top panels), velocity of matter $V_m$ and dark energy $V_{de}$ in the units of Hubble one (middle panels) and gravitational potential $\tilde{\nu}$ (bottom panels) at the central part of spherical halo of cluster scale which is collapsing now. In the top and middle panels solid lines corresponds to matter, dashed lines to dark energy and dotted lines at all panels show the prediction of linear theory.}
\label{collapse}
\end{figure*}

\section*{\sc results and discussion}
\indent In this section we present the results of numerical integrations of the described system of equations. For that we have designed the Fortran-95 program based on open \verb"dverk.f" package for ODEs, in which time depended functions are replaced by scale-factor depended, since $a(t)$ is well defined for given cosmology. The cosmological parameters are taken as follows 
$\Omega_r=4.2\cdot10^{-5},\, \Omega_{de}=0.7,\, \Omega_m=1-\Omega_r-\Omega_{de}$, $H_0=70$ km/s/Mpc. 

The results of numerical integration of system of equations for time evolution of amplitudes of dark energy and dark matter density   and velocity perturbations and potential for different scales and parameters of dark energy are presented in figs. 1 a)-f). In these figures, the solid lines correspond for dark matter perturbations and dashed lines correspond for dark energy ones. The dotted lines at all panels show the predictions of linear theory. For such perturbations non-linearity becomes noticeable already at $a\sim0.1$.

One can note, that in all cases and for all sets of parameters analysed here dark energy is subdominant always: density of its perturbations is few orders lower then of dark matter's ones. The reason of this is the large absolute value of pressure which keeps it from collapse. Large values of effective speed of sound (that is, speed of (also) pressure perturbation distribution) forces dark energy to oscillate after perturbation enters the horizon (figs. $1a$ and $1d$ ). Both velocity and density perturbation of dark energy are smaller during all time of evolution, and the difference is bigger for smaller scales, as can be seen comparing Fig. $1a$, $1b$ and  Fig. $1d$, $1e$.

Another interesting thing is how behaviour of dark energy depends on its parameters. From the figs. a) and c) one can see, that EoS parameter $w_{de}$ slightly changes the character of evolution of inhomogeneities, the initial amplitude of density perturbation is $\sim(1+w_{de})$. At the same time, the effective speed of sound $c_s$ actually defines how fast the perturbation will grow. Comparing figs. a) and b), or fig. d), e) and f) we see, that the closer $c_s^2$ is to zero, the larger amplitude of perturbations is reached at the final stages of evolution. With decrease of effective speed of sound, the pressure gradient of dark energy  decreases counteraction to gravity. The oscillation became smaller and even disappear at some really small values of $c_s$ and the perturbations grow faster, time evolution of velocity and density become more and more similar to the matter's one. This fact is in good concordance with character of dependence on effective speed of sound during accretion of dark energy on compact object, that we studied recently \cite{tsizh1}. As can be seen
from fig. f), when $c_s=0$ the dark energy and dark matter has always the same velocity and the gap between their densities is constant and $\sim(1+w_{de})$ which is velocity factor in the continuity equation for dark energy (\ref{de_cl0}).

These two features, together with the fact, that dark energy background density is smaller than other components' in the past, imply, that dark energy could influence significantly on the process of dark matter hallo collapse only having specific parameters - very small value of effective speed of sound. The current cosmological observational data do not exclude such models of dark energy (\cite{serg2}).

It is also interesting to point out, that predictions of the linear (doted lines) and non-linear theories at smaller scale for density and velocity become distinguishable approximately at the same time (scale factor) as for larger scale, however, for potential, linear and non-linear theories diverge only at late times for smaller scale. The reason for is, that initial perturbations where taken different for these scales, to have approximately the same density perturbation at $a=1$. For the same initial amplitudes the perturbations of smaller scales  would collapse first.   

\section*{\sc acknowledgement}
\indent \indent This work was supported by the projects of Ministry of Education and Science of Ukraine (state registration numbers 0115U003279 and 0113U003059).

\section*{\sc appendix. Newtonian approximation of equations for evolution of dark energy perturbations}

Conservation equation $T^k_{0;k}$ transforms to classic continuity equation for dark energy:
$$\dot{\delta}_{de} + 3\frac{\dot{a}}{a}(c_s^2-w_{de})\delta_{de} + a^{-1}\left((1+c_s^2)\delta'_{de}v + \left[1+w + (1+c_s^2)\delta \right]\left(v' + \frac{2v}{r} \right)\right) = 0 $$
Combination of conservation equations $T^k_{1;k0}$ and $T^k_{0;k}$ transform to classic Euler equation for dark energy:
$$\dot{v} + \frac{v}{a}\left(v'+\frac{2v}{r}\right) + v\frac{\dot{a}}{a}\frac{(1-3c_s^2)(1+w_{de})+(1+c_s^2)(1-3w)\delta_{de}}{1+w_{de}+(1+c_s^2)\delta_{de}}+\frac{\nu'}{2a}+ \frac{c_s^2\delta'_{de}}{a(1+w_{de}+(1+c_s^2)\delta_{de})}=0$$
If we consider the last two equations for dark matter ($c_s^2=w_{de}=0$), we will see that they coincide with well known hydrodynamic equation of collapse \cite{peebles} (equation 9.17)

The non-relativistic approximation of Einstein equation (\ref{ee00}) gives the  Poisson equation for metric function $\nu$ in the coordinates of FRW frame
$$\Delta \nu=8\pi Ga^2 \rho_{cr}^0\left(\Omega_ma^{-3}\delta_m+\Omega_{de}(1+3c_s^2)a^{-3(1+w_{de})}\delta_{de}+2\Omega_ra^{-4}\delta_r\right).$$
The Newtonian gravitational potential $\Phi=\nu/2$.

\end{document}